\documentclass[aps,prb,showpacs,twocolumn,superscriptaddress]{revtex4}
\usepackage[english]{babel}
\usepackage{graphicx}
\usepackage{color}
\usepackage[utf8]{inputenc}
\usepackage{pstricks,pst-grad,color}
\usepackage{graphicx,pifont,amssymb}
\usepackage{amsmath,amssymb}

\begin{document}


\title{Kondo effect in {\it f}-electron superlattices}



\author{Robert Peters}
\email[]{peters@scphys.kyoto-u.ac.jp}
\affiliation{Department of Physics, Kyoto University, Kyoto 606-8502,
  Japan}

\author{Yasuhiro Tada}
\affiliation{Institute for Solid State Physics, The University of
  Tokyo, Kashiwa 277-8581, Japan}

\author{Norio Kawakami}
\affiliation{Department of Physics, Kyoto University, Kyoto 606-8502,  Japan}

\date{\today}

\begin{abstract}
We demonstrate the importance of the Kondo effect in artificially
created {\it f}-electron superlattices.
We show that the Kondo effect does not only change the density of
states of the {\it f}-electron layers, but is also the cause of pronounced
resonances at the Fermi energy in the density of states of the 
non-interacting layers in the superlattice, 
which are between the {\it f}-electron layers.  Remarkably, these
resonances strongly depend on the structure 
of the superlattice; due to interference, the density of states at the
Fermi energy can be strongly enhanced or even shows no changes at
all. Furthermore, we show
that by inserting the Kondo lattice layer into a three-dimensional
(3D) metal, the 
gap of the Kondo insulating state changes from a full gap to a pseudo
gap with quadratically vanishing spectral weight around the Fermi energy.
Due to the formation of the Kondo insulating state in the
{\it f}-electron layer, the superlattice becomes
strongly anisotropic below 
the Kondo temperature. We prove this by calculating the in-plane and
the out-of-plane conductivity of the superlattice. 
\end{abstract}


\maketitle

\section{Introduction}

Recently, remarkable progress has been made in the creation of
artificial superlattices 
including {\it f}-electron materials.\cite{Shishido2010,Mizukami2012,Goh2012}
These superlattices consist of a periodic arrangement of
two-dimensional {\it f}-electron layers, which are inserted into a
different material. In the experiments, these superlattices show
some intriguing  
magnetic and superconducting properties, which are
tunable by changing the superlattice structure.
For example, the N\'eel temperature in
CeIn$_3(n)$/LaIn$_3$(4) superlattices\cite{Shishido2010}
decreases to zero when the Cerium layer thickness $n$ is reduced to $n=2$,
which is accompanied by a linear temperature dependence of the
resistivity. In other experiments, using the heavy fermion
CeCoIn$_5$ and the conventional metal YbCoIn$_5$, a two-dimensional
strong coupling superconducting state has been
observed.\cite{Mizukami2012} Also in 
these experiments, the transition 
temperature and the magnetic field dependence are strongly affected by the
structure of the superlattice.
The ability to create new materials with properties that
depend on the superlattice structure may provide the possibility to
construct new functional devices. 

In {\it f}-electron materials, the competition or cooperation of two
mechanisms mainly determines the properties of the material, i.e. the
Kondo effect and the RKKY interaction.\cite{doniach77} The quantum
criticality arising 
when both effects are equally strong, causes non-Fermi
liquid behavior and unconventional superconductivity.\cite{coleman2007,coleman2005,gegenwart2008} In order to
understand these intriguing effects in the context of these superlattices,
it is essential to understand how 
both mechanisms, which are the cause of most of the phenomena,
are influenced by the superlattice structure. Therefore, in this paper
we will study 
the Kondo effect in the paramagnetic state of {\it f}-electron
superlattices.
We will clarify how the Kondo effect influences the {\it f}-electron
layers and in what
way the Kondo effect is observable in the non-interacting layers, which are
in between the {\it f}-electron layers. 
Another interesting question which
we will address is the influence of the superlattice structure on the
low energy behavior of a Kondo lattice: What happens to the full gap
of the isolated Kondo lattice 
layer, when it is coupled to metallic layers. 

In order to answer these questions, an essential point is the proper
treatment of the 
superlattice structure. Therefore, we simulate systems which include
two-dimensional (2D) {\it f}-electron layers and 2D non-interacting 
layers. The number and order of these layers can be varied, so that a
variety of different superlattices can be studied. We
are thus able to analyze these {\it f}-electron superlattices and the Kondo
effect within these systems taking the structure of the superlattice fully
into account. We find that the hybridization gap in the {\it f}-electron
layers changes into a pseudo gap as soon as different layers are
coupled to each other. However, the Kondo effect in the {\it
  f}-electron layers also strongly affects the density of states
of the non-interacting layers, corresponding to Yb- or La-layers in
the above-mentioned experiments. We find that resonances,
whose shapes 
depend on the superlattice structure and are influenced by
interference from different {\it f}-electron layers, appear at the
Fermi energy.  These resonances have strong influence on the local
properties of these superlattices and might be observable by local
probes in experiments.

The remainder of this paper is organized as follows: In the next section,
we explain in detail the model and the method which we use in our
study. This is followed by a section where we study a single {\it
  f}-electron layer within a 3D lattice. This kind of study will give
insights into the physics which can be expected in the
superlattice. Finally, we present results for the paramagnetic state
in {\it f}-electron superlattices, analyzing a wide range of different
superlattice structures. A summary will conclude this paper.

\section{Model and Method}
\begin{figure}[t]
\includegraphics[width=\linewidth]{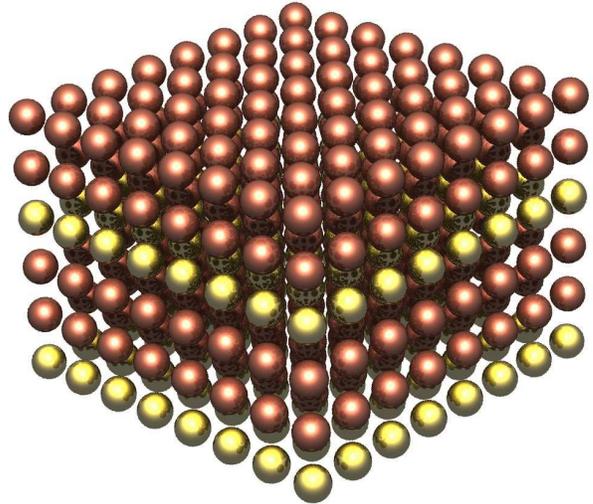}
\caption{(Color online) Schematic picture of a (1,2)-superlattice,
  consisting of a 
  periodic arrangement of one {\it f}-electron layer and two
  non-interacting layers. 
 \label{superlatt}}
\end{figure}

We simulate a
three dimensional system, 
consisting of a periodic arrangement of {\it f}-electron- and
non-interacting layers, see Fig. \ref{superlatt}. The Hamiltonian of the
superlattice can be 
written as a sum of terms 
describing 2D non-interacting layers, $H_{\text{NIL}}$,
{\it f}-electron layers described by a 2D Kondo lattice
model,\cite{doniach77,lacroix1979,fazekas1991}  
$H_{\text{KLL}}$, and a hopping, $H_{\text{Inter}}$, which connects adjacent layers with
each other: 
\begin{eqnarray}
H&=&H_{\text{NIL}}+H_{\text{KLL}}+H_{\text{Inter}},\label{Ham}\\
H_{\text{NIL}}&=&t_{\text{NIL}}\sum_{<i,j>\sigma}c^\dagger_{iz\sigma}c_{jz\sigma},\nonumber\\
H_{\text{KLL}}&=&t_{\text{KLL}}\sum_{<i,j>\sigma}c^\dagger_{iz\sigma}c_{jz\sigma}+J\sum_i\vec{S}_{iz}\cdot\vec{s}_{iz},\nonumber\\
H_{\text{Inter}}&=&t_{z}\sum_i\sum_{<z_1,z_2>\sigma}c^\dagger_{iz_1,\sigma}c_{iz_2\sigma}.\nonumber
\end{eqnarray}
In this paper we describe each layer as a 2D square lattice for which
$i$ and $j$ are indices within the same layer. The layer index is given by $z$.
Thus, the operator $c_{iz\sigma}^\dagger$ creates an electron at lattice
site $i$ in layer $z$ in spin direction $\sigma$. Besides a nearest
neighbor hopping within the 3D superlattice, there is an additional
spin-spin interaction, $J\sum_i\vec{S}_{iz}\cdot\vec{s}_{iz}$, between
the spin of the electrons and the localized spins in the Kondo
lattice layers. 
If not otherwise stated, all hopping constants are equal
$t_{\text{NIL}}=t_{\text{KLL}}=t_z=t$. We take the 2D
half-bandwidth $W_2=4t$ as unit of energy. Furthermore, we assume
throughout this paper an antiferromagnetic coupling, $J>0$, between
the conduction electrons and the localized spins, which is appropriate to
describe heavy fermion materials.
Without any coupling to the
spins in the Kondo lattice model, the system corresponds to
non-interacting electrons in a 3D cubic lattice. In the rest of the
paper, we will characterize 
the superlattice structure as $(N,M)$, corresponding to
KLL$_N$/NIL$_M$ ($N$: Kondo lattice layers (KLL), $M$: non-interacting
layers (NIL)). While we assume that the system is homogeneous within
each plane, i.e. within the same plane all lattice sites are equal, we
use open boundary conditions perpendicular to the layered structure including 
up to $40$ layers in the superlattice. Throughout the paper, we focus
on physical properties of the center-layers, which do not show any signs
of boundary effects.

For solving this model, we use a combination of the inhomogeneous dynamical
mean field theory (IDMFT)
and the numerical renormalization group
(NRG).\cite{wilson1975,bulla2008} 
IDMFT originates in the dynamical mean field theory (DMFT).\cite{georges1996}
While non-local contributions to the self-energy 
are neglected, local fluctuations
and the layer dependence of the self-energy are fully taken into
account within IDMFT. The Kondo effect as well as the heavy fermion
state are well described, and the phase diagram of the Kondo lattice model
within DMFT has been investigated by a number of authors
before.\cite{jarrell1993,sun1993,rozenberg1995,sun2005,peters2007,leo2008,otsuki2009,hoshino2010,bodensiek2011,peters2012,peters2013}

In the original DMFT approach, one is interested in a homogeneous
lattice, so that each site can be described by the same self-energy.  
Because in the superlattice the system is inhomogeneous, i.e. there
are different 
kinds of layers with different physical properties, different lattice
sites cannot be approximated with the same self-energy. Instead, we
have to take into account the structure of the superlattice and
calculate a site dependent self-energy. In the case of the {\it
  f}-electron superlattice, Eq. (\ref{Ham}), the self-energy for the
NILs vanishes. IDMFT works as follows:

{\it a)} We start with a guessed self-energy, which can also be set
to zero.

{\it b)} Using this self-energy, we calculate the site dependent local
Green's functions of the superlattice. Because our system is
translation invariant within each layer, we use Fourier
transformation for lattice sites within the same layer. The local
Green's function of layer $z$, $G_{z}(\omega+i\eta)$, thus reads
\begin{eqnarray}\label{local_Green}
G_{z}(\omega+i\eta)&=&\frac{1}{4\pi^2}\int dk_xdk_y \Bigl( (\omega+i\eta)\mathbb{I}\\&&-H(k_x,k_y) -\Sigma(\omega+i\eta)\Bigr)^{-1}_{z},\nonumber
\end{eqnarray}
where $H(k_x,k_y)$ is a matrix corresponding to the
one-particle Hamiltonian of the 
system, i.e. Eq. (\ref{Ham}) with $J=0$, in which for the in-plane hopping
Fourier transformation has been used, resulting in
$H_{\text{NIL}}=2t(\cos(k_x)+\cos(k_y))$. Eq. (\ref{local_Green}) involves a
matrix inversion of dimension corresponding to the number of layers in
the system.
Interaction effects are taken into account by the self-energy
$\Sigma(\omega+i\eta)$, which depends on the layer.
Furthermore, we take $\eta$ small
enough, so that the resulting Kondo gap can be described. A typical
value in our calculations is $\eta=10^{-5}$.

{\it c)} We now relate the calculated local Green's function of each
layer with the Green's function of the corresponding quantum impurity
model using the same self-energy. The impurity Green's function can be written as\cite{Hewson1997}
\begin{equation}
G_{\text{IMP}}(\omega+i\eta)=\frac{1}{\omega+i\eta-\Delta(\omega+i\eta)-\Sigma(\omega+i\eta)}.
\end{equation}
$\Delta(\omega+i\eta)$ is the hybridization between the impurity site and the
conduction electron bath. Due to the locality of the self-energy, we can
assume that the calculated lattice Green's function is equal to the
impurity Green's function, from which we can determine the hybridization
function
\begin{equation}
\Delta(\omega+i\eta)=\omega+i\eta-G_{z}^{-1}(\omega+i\eta)-\Sigma(\omega+i\eta).
\end{equation}
Within IDMFT, the hybridization function, $\Delta(\omega+i\eta)$, depends on
the layer. 

{\it d)} The calculated hybridization functions define for each layer
a quantum impurity model. We solve for each layer the
corresponding impurity model and calculate its self-energy. 

{\it e)} We iterate this procedure by calculating new local lattice
Green's functions, see {\it b)}. If the self-energies of all layers between two
iterations do not change, the IDMFT solution is found.

Besides the calculation of local Green's functions, the
time consuming step is the calculation of the self-energy for each layer.
For this purpose we use the NRG.\cite{bulla2008}
The NRG is able to solve these quantum impurity models 
for a wide range 
of temperatures and is able to resolve exponentially small energy
scales like the Kondo resonance using the complete Fock space
algorithm.\cite{peters2006,weichselbaum2007} 
Throughout this paper, we use a discretization parameter $\Lambda=2$ and
keep $N=1000$ states within each NRG iteration.\cite{bulla2008}

\section{A single 2D Kondo lattice layer within a 3D metallic host}
\begin{figure}[t]
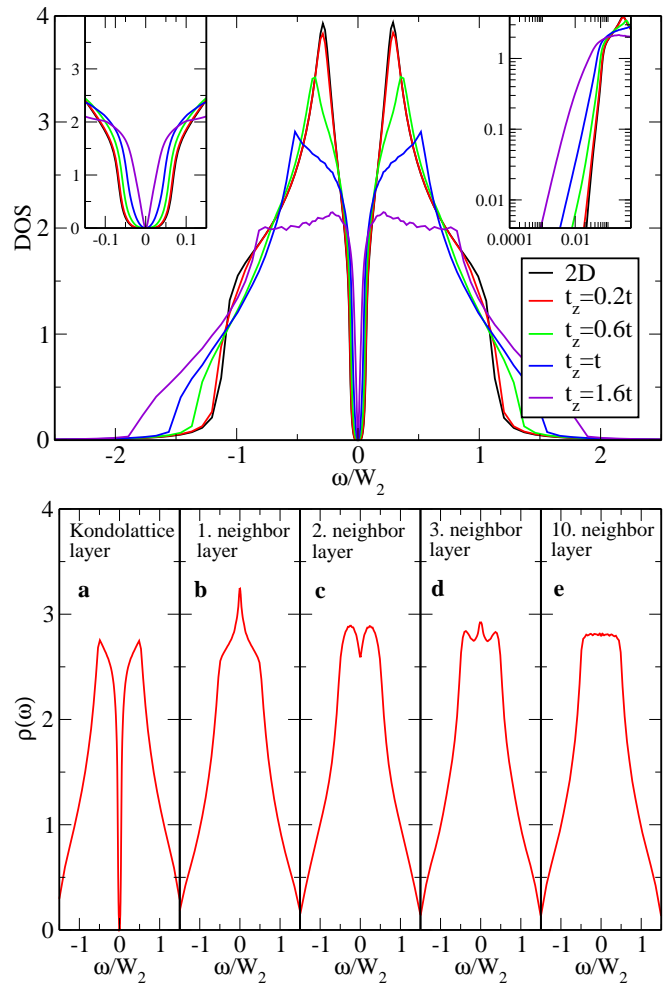

\includegraphics[width=\linewidth]{fig2a.eps}
\includegraphics[width=\linewidth]{fig2b.eps}
\caption{(Color online) Results of a single KLL, $J=0.5W_2$, within a 3D
  non-interacting host. Upper panel: DOS of the KLL for different
 hopping amplitudes, $t_z$, between the KLL and the NIL. The insets
  show magnifications around the Fermi energy. Lower panel: DOS of
  different layers. a) DOS of the KLL, b) 
nearest neighbor layer, c) next nearest neighbor layer, d)
third-nearest neighbor layer, e) 10 layers away.
 \label{single_Green}}
\end{figure}
In this section, we will first analyze a single KLL within a 3D
metal. This will give insights into changes occurring due to the
insertion of a KLL into a metallic host.

The paramagnetic Kondo lattice model at half filling exhibits a gap in
the density of states (DOS) at the Fermi energy. The system is in the Kondo insulating
state. 
The first question, we want to clarify is, how the DOS of the
Kondo lattice is modified, if it is inserted into a 3D metal. 
What happens to the full gap, if it is coupled to a metallic layer?
The electrons of the metallic layer may tunnel
into the KLL, thus, modifying the DOS. 

We show the DOS of a single KLL
within a 3D metal for different inter-layer hopping amplitudes, $t_z$, 
in the upper panel of Fig. \ref{single_Green}. The insets show
magnifications around the Fermi energy. 
First, when increasing the inter-layer hopping amplitude, the gap
width decreases. Second, the double-logarithmic inset clearly
shows that the spectral weight vanishes as a power law for small
frequencies, if the KLL is coupled to a metallic layer.
The determined exponent is approximately two and seems to be
independent of the hopping amplitude. The prefactor in front of the
quadratic term, 
however, does depend on the hopping amplitude and increases with
increasing hopping.
An arbitrary small coupling, $\vert t_z\vert>0$, between
  different layers is sufficient to transform the full gap of an
  isolated KLL into a pseudo gap.
We note here that even for the isolated KLL the
edges of the gap are broadened due to interaction effects.

Another noticeable change occurs in the DOS away from the Fermi
energy. Increasing the coupling between the KLL and the NILs, 
the two-peak structure, which is a remnant of the van Hove singularity
in the two-dimensional square lattice, is flattened and vanishes. 
The hopping between
different layers creates a 3D structure for the 
conduction electrons, and thus destroys these singularities. 

To get some more insights into the formation of the pseudo gap, we
look at a non-interacting analog, 
consisting of two 
layers, which are coupled via hopping $t$. 
While one layer is a non-interacting metallic layer, the other layer
is described by a non-interacting periodic Anderson model.
 In the periodic Anderson
model each lattice site is hybridized via a hopping $V$ to a localized
{\it f}-electron level. The model can be written as
\begin{displaymath}
H=\sum_{ij,\sigma}t_{ij}c_{i\sigma}^\dagger
c_{j\sigma}+V\left(c_{i\sigma}^\dagger f_{i\sigma}+f_{i\sigma}^\dagger c_{i\sigma}\right),
\end{displaymath}
where $f_{i\sigma}^\dagger$ creates an {\it f}-electron. Due to this
hybridization, a gap opens around the Fermi energy.
In the case of the
non-interacting periodic Anderson model, the spectral weight at the
edges of the gap jumps from a finite value to zero. For the
Kondo lattice model, this jump is smeared out due to
two-particle interactions between the conduction electrons and the
localized spins. 
If the layer described by the periodic Anderson model is coupled to
the metallic layer via a hopping $t$, 
the DOS can be written as
\begin{eqnarray*}
\rho(\omega)&=&\int_{-D}^D d\epsilon \frac{1}{\omega+i0-\epsilon-\frac{V^2}{\omega+i0}-\frac{t^2}{\omega+i0-\epsilon}},
\end{eqnarray*}
where we have assumed for simplicity a constant DOS within
each layer. The result of this integral shows that the gap of the {\it
  f}-electron layer is changed by the coupling to a metallic layer
from a full gap to a pseudo gap. 
The existence of the gap itself is unchanged. The local hybridization
within the {\it f}-electron layer is sufficient to open a gap.
However, inside the original gap, we find that the spectral weight
vanishes quadratically.
The change between a full gap and the pseudo gap occurs at arbitrary
small hopping between these two layers and
happens in the same way in the interacting Kondo lattice model and the
non-interacting periodic Anderson model. The exponent of the power
law, which determines how the spectral weight vanishes around the
Fermi energy, seems to be equal in both models.

The coupling between the KLL and the NILs does not only
modify the DOS of the KLL. It has also profound influence on the DOS
of the neighboring NILS.
In the lower panel of Fig. \ref{single_Green}, we show the DOS of
different layers in a   
system, where only one KLL is inserted into a 3D metal.
The hopping is isotropic, $t_z=t$.
The DOS of the KLL (panel a) shows the above-described pseudo gap around
the Fermi energy, $\omega=0$. 
The DOS of the neighboring layer (panel b), which is a non-interacting
layer, shows a modification from the usual 3D 
DOS; there
is a peak at the Fermi energy, whose width corresponds to the gap
width of the KLL. The next nearest layer, on the other
hand, shows a dip at the Fermi energy. 
The peak or gap at the Fermi energy alternates with
increasing distance from the KLL, showing typical $2k_F$
oscillations. 
The width of these structures at
the Fermi energy is completely determined by the gap width of the
KLL, thus, they are supposed to be related to the Kondo
effect (Further evidence that the Kondo
effect plays a significant role is shown later). However, the amplitude of these structures
decreases quickly, roughly as $1/d^\kappa$
  ($\kappa\approx 1.1$, $d$: distance between
  the layers), when going further away from the KLL. These
  oscillations occur in a  similar way as Friedel oscillations around
  impurities in metals.
Panel
e) in Fig. \ref{single_Green} shows the DOS of an atom 10 layers
away from the KLL, where the DOS resembles the unperturbed 3D DOS.

\section{{\it \lowercase{f}}-electron superlattices}
\begin{figure}[t]
\includegraphics[width=\linewidth]{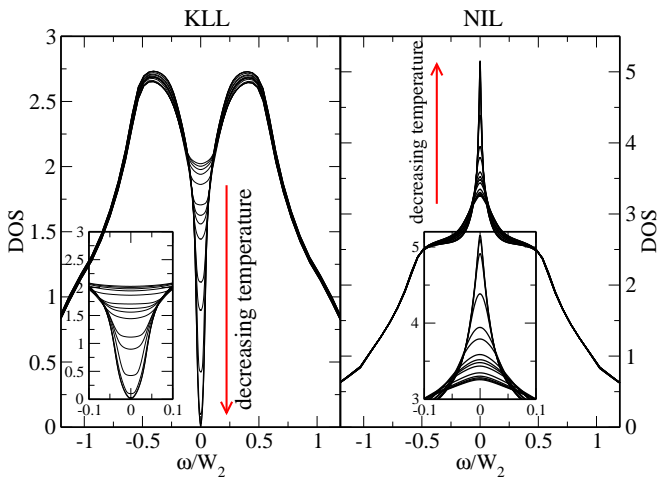}
\caption{(Color online) DOS of a (1,1)-superlattice for different
  temperatures. The insets show magnifications around the Fermi energy.
 \label{GreenAB}}
\end{figure}

\begin{figure}[t]
\includegraphics[width=\linewidth]{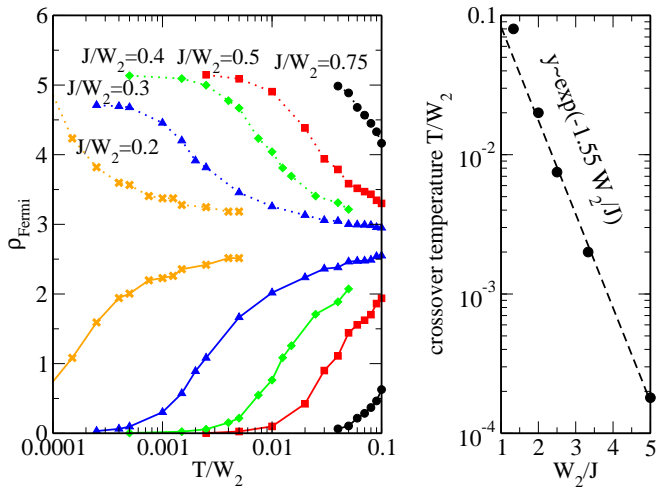}
\caption{(Color online)  DOS at the Fermi energy and crossover
  temperature for the (1,1)-superlattice. 
Left: DOS at the Fermi energy, $\rho_{Fermi}$ for different
temperatures and interaction strengths. The lower (upper) lines
correspond to the KLL (NIL). Right:
crossover temperature over inverse interaction strength, proving the
exponential dependence.
 \label{TkAB}}
\end{figure}

Next, we will study the Kondo effect in {\it f}-electron
superlattices, i.e. in a periodic arrangement of 2D KLLs
and NILs. Figs. \ref{GreenAB} and \ref{TkAB} show results for  
a (1,1)-superlattice, where the system is composed alternatively of
KLLs and NILs. The behavior of the DOS in this superlattice is similar
to the alternating peak/dip 
structure at the Fermi energy of the system with a single KLL in 3D (Fig. \ref{single_Green}). 
The KLL shows again the formation of a pseudo gap, and the DOS of the nearest neighbor NIL shows a peak
at the Fermi energy. 
However, if we compare the amplitude of the peak at the Fermi energy
of the NIL to the 
DOS of the unperturbed 3D metal, the DOS 
is enhanced nearly by a factor of two.
Because each NIL is sandwiched by two KLLs in this (1,1)-superlattice
structure, the Kondo effect of both KLLs enhances the DOS at 
the Fermi energy of the NIL. One may refer to this as
``constructive interference''. 
In Fig. \ref{GreenAB}, we also include curves for different
temperatures, showing how the gap in the KLL and the
peak in the NIL are built up with decreasing temperature.
At high temperatures, $T\gg T_K$, where $T_K$ is the Kondo
temperature of the system, one can clearly see a dip formation in the
DOS of the KLL and a bump in the DOS of the NIL. These structures are
formed at temperatures $T\approx J$. When lowering the temperature
below the Kondo temperature, the dip becomes more pronounced and
eventually forms the Kondo gap, while the peak in the DOS
of the NIL increases at the same time.
A closer analysis of the influence of the temperature on the DOS at
the Fermi energy 
for the (1,1)-superlattice is
presented in Fig. \ref{TkAB}. The left panel shows the DOS at the
Fermi energy of both layers for different temperatures and different
coupling strengths. For all coupling strengths, there is a
crossover from the high temperature phase, where both layers have
nearly equal weight at the Fermi energy, to the low temperature phase
consisting of Kondo gapped layers and NILs, which have enhanced weight
at the Fermi energy.
The crossover temperature, which is approximately equal to the width
of the gap of the KLL or the width of the peak of the NIL at the Fermi energy,
decreases with 
decreasing coupling strength. We show these crossover temperatures
over the inverse 
coupling strength in the right panel of Fig. \ref{TkAB}, proving the
exponential dependence on the coupling strength $J$. Because the peak
in the DOS of the NIL shows the same temperature dependence as the gap
of the KLL, we can 
state that both resonances are induced by the Kondo effect of the {\it
  f}-electrons. The qualitative behavior of the system
  is independent 
  of the coupling strength $J$. The coupling strength only determines the Kondo
  temperature below which the resonances at the Fermi energy can be
  observed, as long as a paramagnetic state is formed.

\begin{figure}[t]
\includegraphics[width=\linewidth]{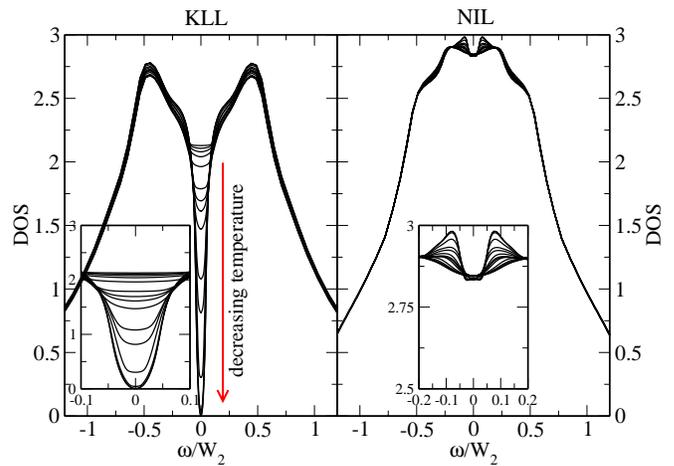}
\caption{(Color online) DOS of a (1,2)-superlattice for different
  temperatures. Due to symmetry, both NILs are identical. 
 \label{GreenAAB}}
\end{figure}
\begin{figure}[t]
\includegraphics[width=\linewidth]{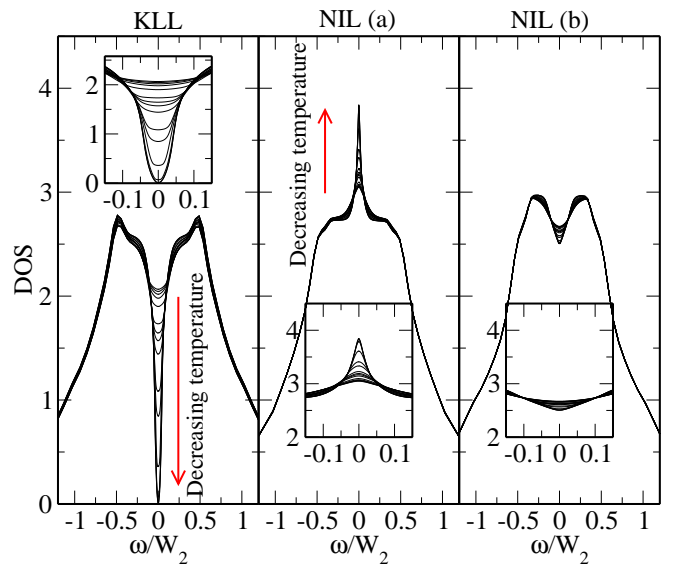}
\caption{(Color online) DOS of a (1,3)-superlattice for different
  temperatures. 
 \label{GreenAAAB}}
\end{figure}
If the structure of the superlattice is changed, the resonances
at the Fermi energy are also altered.
In Figs. \ref{GreenAAB} and \ref{GreenAAAB}, we show the temperature
dependent DOS of the (1,2)- and (1,3)-superlattices, which consist of
two or three NILs in the unit cell, respectively.
For both configurations, the KLLs form the pseudo gap at the Fermi energy
at low temperatures. This behavior is not 
changed by tuning the 
structure of the superlattice. Independent of the superlattice
structure, the KLLs 
always form Kondo gapped states at half filling and zero
temperature. The NILs, on the 
other hand, show different behavior depending on the superlattice structure.
For the (1,2)-superlattice in Fig. \ref{GreenAAB}, both non-interacting
layers exhibit identical DOS, because of the symmetry of the system.
Surprisingly, the DOS at the Fermi
energy of the NILs remains unchanged for all
temperatures. This is in contrast to the (1,1)-superlattice
configuration shown in Fig. \ref{GreenAB}. In the (1,2)-superlattice,
we observe only 
small temperature-dependent changes away from the Fermi
energy, $\omega\approx\pm 0.1W_2$ in Fig. \ref{GreenAAB}, but there is no
peak or dip at the Fermi energy.
This can be explained by the peak/dip oscillations seen in the
DOS of a single KLL in 3D, shown in Fig. \ref{single_Green}. The nearest
neighbor NIL exhibits a  
peak at the Fermi energy, while the next nearest layer shows
a dip. In the (1,2)-superlattice, there is thus a ``destructive
interference''  of both effects. Each NIL is nearest
neighbor as well as next nearest neighbor to a KLL.
This results in an
unaltered DOS of the NIL at the Fermi energy. 
Therefore,  the influence of the Kondo effect on the DOS at the Fermi
energy is only visible in the KLLs for this configuration.
The situation is again changed for a (1,3)-superlattice, shown in
Fig. \ref{GreenAAAB}. 
This structure favors a
constructive interference. Thus, while the KLL forms the Kondo
gap, the NIL which is nearest to the KLL shows a peak at the
Fermi energy. The peak height is reduced compared to the
(1,1)-superlattice (see Fig. \ref{GreenAB}). The second NIL, which is in
the middle of three NILs,
shows a small dip at the Fermi energy. This agrees with the peak/dip
oscillations found in the results of a single KLL in 3D.
\begin{figure}[t]
\includegraphics[width=\linewidth]{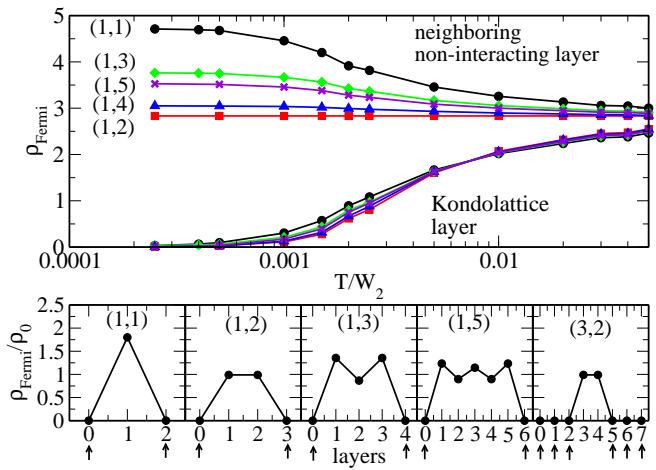}
\caption{(Color online) Upper panel: DOS at the Fermi energy over
  temperature for 
  $J/W_2=0.3$, and for variety of superlattice configurations. We only
  show the Kondo lattice layer and the nearest neighbor NIL.
Lower panel: Each panel shows the normalized DOS at the Fermi energy
for all different layers in the superlattice. The arrows at the
  layer index denote the KLLs. 
 \label{diff_fermi}}
\end{figure}

\begin{figure}[t]
\includegraphics[width=\linewidth]{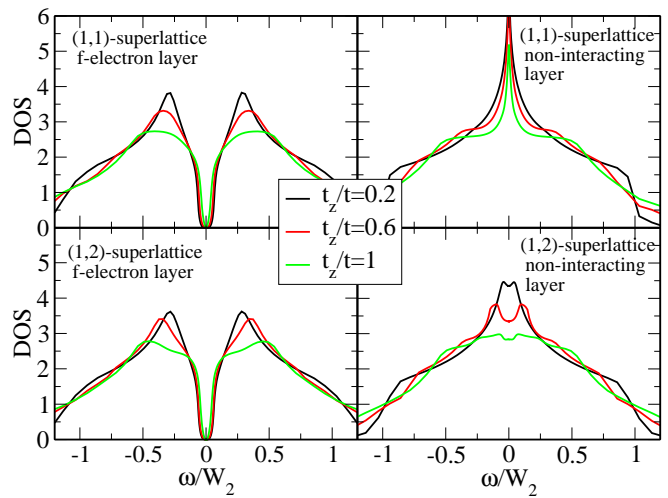}
\caption{(Color online) 
 \label{diff_tz_superlatt} Spectral functions at $T=0$ for the $(1,1)$-
and $(1,2)$-superlattice for different inter-layer couplings, $t_z$.}
\end{figure}

Finally, we compare the DOS at the Fermi energy for a variety of superlattice
configurations in Fig. \ref{diff_fermi}.
The  upper panel in Fig. \ref{diff_fermi} shows
the temperature
dependence of the DOS at the Fermi energy for the KLL and the
nearest neighboring NIL for 
fixed interaction strength, $J/W_2=0.3$. 
On the one hand, any KLL forms for any superlattice configuration the Kondo
gap at the Fermi energy. 
On the other hand, we observe that the
amplitude of the peak in the neighboring NIL strongly
depends on the superlattice structure, due to interference between
the Kondo effect of different {\it f}-electron layers.
Because of the
above-described constructive/destructive interference, the peak height
changes non-monotonically when inserting additional NILs, and shows
even-odd oscillations.  
Looking
only at an odd number of NILs, which always results in a constructive
interference for the nearest neighbor NIL, the peak height is
decreased when inserting  
new NILs. The maximum peak height can be observed for the (1,1)-superlattice.
On the other hand, for an even number of NILs, the peak height is
increased when inserting new layers. 
For the (1,2)-superlattice, there is complete destructive interference and
consequently no peak emerges at the Fermi energy. Furthermore, the
crossover temperature seems to be slightly dependent on the superlattice
structure. This is best visible in the temperature dependence of the
KLLs in Fig. \ref{diff_fermi}. 
Note that the crossover temperatures of different superlattices are changed according to the peak heights in the DOS of the
NIL at the Fermi energy. 
 The lowest crossover temperature can be observed for the
(1,1)-superlattice, and the highest crossover temperature can be
observed for the (1,2)-superlattice. The difference between the (1,1)-
and the (1,2)-superlattice is about $20\%$, which should be detectable
in experiments.

Up to now, we have focused on periodic configurations of a
single KLL and a certain number of NILs. Thus,
there are never two KLLs touching each other. The
reason for our choice is that the physics, which has been so far explained,
does not change 
when adding additional KLLs. There are only small
changes in the DOS away from the
Fermi energy. Any KLL forms a pseudo gap within the superlattice
structure, for which the spectral weight around the Fermi energy
vanishes quadratically. Only the prefactor in front of the quadratic term
does depend on the superlattice structure.  
Furthermore, the resonances observed at the Fermi energy of the NILs,
are only determined by the number of the NILs.

In the lower panel of Fig. \ref{diff_fermi}, we show the changes of
the DOS at the Fermi 
energy for all layers in different superlattices at zero temperature. 
As in the results for the single KLL embedded in 3D, we observe
oscillations of peaks and dips at the Fermi energy of the NILs.
Configurations having two NILs are special, as described above. We observe
that for these superlattices  ((1,2)- and (3,2)-superlattice in
Fig. \ref{diff_fermi}), the
DOS at the Fermi energy is not changed and is independent of the number of
KLLs. 
For the other superlattice configurations, we see that the amplitude of
the resonances decreases 
with inserting new NILs. However, even for the (1,5)-superlattice, where
there are five NILs between the KLLs, the enhancement at the Fermi energy
is still around 20\%, for the (1,3)-superlattice even 40\%.
This shows that the Kondo effect in the {\it f}-electron layers has
substantial influence on all NILs and should not be dismissed, even if
there are several layers between the {\it f}-electron layers.
In the experiments,\cite{Shishido2010,Mizukami2012,Goh2012} there are
$4$ or $5$ NILs in between the {\it f}-electron layers.
Our results imply that the Kondo effect penetrates through the NILs at
low temperatures, which will lead to a coupling of the separated {\it
  f}-electron layers.

An open question, which we want to answer is, how the 
superlattice changes into a 2D system, when the inter-layer
hopping is decreased. We show this process exemplary
for the $(1,1)$- and $(1,2)$-superlattices in
Fig. \ref{diff_tz_superlatt}. With decreasing the inter-layer hopping,
$t_z$, each layer behaves more and more like a 2D square lattice without
qualitatively altering the behavior at the Fermi energy as
described above. 
The DOS of the {\it
  f}-electron layer exhibits a hybridization gap at the Fermi
energy. The gap width becomes
slightly increased when decoupling the layers, because the gap width
in the 2D Kondo lattice is larger than in the 3D Kondo lattice. The DOS
of the NILs, on the other hand, is expected to
evolve into the DOS of a 2D square lattice with van Hove singularity
at the Fermi energy. This can be seen in the right panels of
Fig. \ref{diff_tz_superlatt}. For the $(1,1)$-superlattice, there is a
peak in the DOS of the NIL for all calculated values of the
inter-layer hopping. However, while in the DOS of a nearly decoupled
NIL the peak 
corresponds to the van Hove singularity, the peak in the coupled
system, $t_z/t=1$, originates in the Kondo effect, which strongly
depends on the
temperature, as has been shown in Fig. \ref{GreenAB}. The spectral
weight around the Fermi energy also increases for the
$(1,2)$-superlattice.
The two peak structure close to the Fermi energy remains observable
but begins to merge
into the van Hove singularity when decreasing the inter-layer hopping.

\begin{figure}[t]
\includegraphics[width=\linewidth]{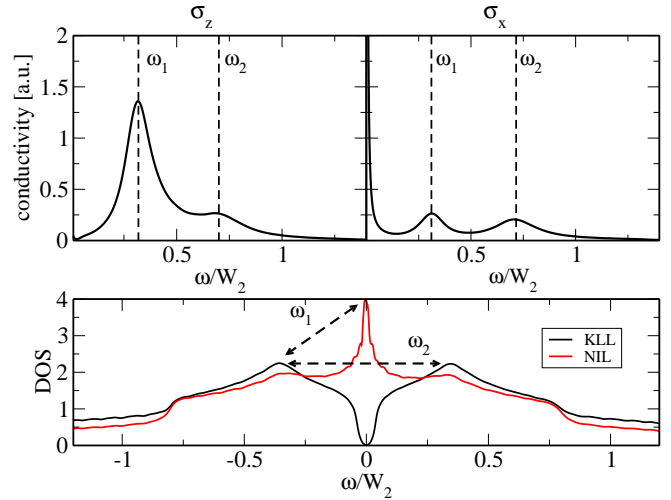}
\caption{(Color online) Real part of the conductivity within the
  layers, $\sigma_x$, and 
  perpendicular to the superlattice structure, $\sigma_z$, for
  $J/W_2=0.5$.  The lower panel shows the excitations in the local DOS
  corresponding to the peaks in the conductivity.
 \label{conAB}}
\end{figure}
The formation of a gap at the Fermi energy in the DOS of the KLLs
and a peak in the DOS of the NILs, results in a highly anisotropic system
at low temperatures.
In order to elucidate this point further, we show the real part of the
frequency dependent 
conductivity of the (1,1)-superlattice for $J/W_2=0.5$ at zero
temperature in Fig. \ref{conAB}. 
Writing the one-particle part of the Hamiltonian, Eq. (\ref{Ham}), in momentum
space, the current operator is given as\cite{pruschke2003}
\begin{equation}
\vec{j}=e\sum_\sigma\sum_{\vec{k}}c_{\vec{k}\sigma}^\dagger \vec{\mathbf{v}}(\vec{k})c_{\vec{k}\sigma},
\end{equation}
where $\vec{\mathbf{v}}(\vec{k})=\frac{\partial
  H(k_x,k_y,k_z)}{\partial \vec{k}}$, and $e$ is the elementary
charge. For calculating the conductivity, we use periodic boundary
conditions in all directions.
The conductivity is then given by the Kubo formula via the
dynamical current-current 
correlation function, where the current is taken in direction $\mu$, and reads,
\begin{eqnarray}
\sigma_\mu(\omega)&=&\frac{1}{i\omega} \langle\langle
j_\mu,j_\mu\rangle\rangle_{\omega+i\delta}.
\end{eqnarray}
Due to the locality of the self-energy within DMFT, the resulting two-particle Green's functions can be
written as products of one-particle Green's functions.\cite{pruschke2003}

In Fig. \ref{conAB}, we show the out-of-plane conductivity in the
upper-left panel 
and the in-plane conductivity in the upper-right panel. The lower
panel shows the local DOS of both layers. 
The anisotropy of the system is clearly visible in the conductivities
at low frequencies. 
The in-plane conductivity, $\sigma_x$, exhibits a strong peak for
$\omega/W_2<0.05$. This peak signalizes metallic behavior
parallel to the planes. Analyzing the different
components in the conductivity, we observe that this peak comes mainly
from the NILs, 
having strong weight in the local DOS at the Fermi energy. On
the other hand, 
the out-of-plane conductivity, $\sigma_z$, shows a gap at low
frequencies. Thus, perpendicular to the layered structure the {\it
  f}-electron superlattice is insulating. 
At the Kondo temperature, the superlattice becomes strongly
anisotropic, showing metallic behavior within the planes and
insulating behavior perpendicular to the planes. 
Analyzing the conductivity at higher frequencies, we observe two distinct
peaks. These two frequencies can be easily identified 
within the DOS of both layers (lower panel of Fig. \ref{conAB}); one
corresponding to the excitation energies between the lower band and
the central peak, and the other corresponding to the excitations
between the lower and upper bands.  
Even if the superlattice structure is altered, the conductivity remains
qualitatively unchanged. The superlattice still behaves as an
anisotropic metal at low 
temperatures, i.e. perpendicular to the planes the superlattice is
insulating and parallel to the planes metallic.
The number and position of these
peaks, however, depend on the superlattice structure. 
At large frequencies, 
additional peaks are visible for different superlattice structures,
arising from excitations between 
different layers. The temperature-dependent
  conductivity, especially the influence of the superlattice on it, is
  analyzed in detail in reference.\cite{Tada2013}

\begin{figure}[t]
\includegraphics[width=\linewidth]{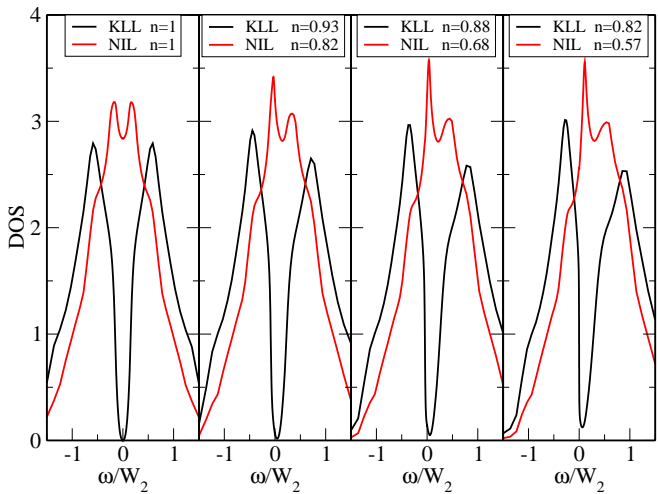}
\caption{(Color online) DOS for the (1,2) superlattice in a doped system   $J/W_2=0.65$.
 \label{dopeAAB}}
\end{figure}
All results, which have been shown until now, have been for half filled
lattices, for which each lattice site is occupied in average with one
electron and the KLLs form the Kondo gap at the Fermi energy. If the
chemical potential is changed, the system is doped 
away from half filling.
 However, the main
statements about the Kondo effect, which we have made, remain
valid. In Fig. \ref{dopeAAB}, we show the local DOS of a
(1,2)-superlattice 
for different chemical potentials. 
The structures (peaks and dips) in the DOS of the NILs are shifted.
By
doping the system, one of the side peaks in the DOS of the NIL is
shifted towards 
the Fermi energy. At the same time the gap in the KLLs is shifted away
from the Fermi energy.
Thus, even for the doped system the Kondo effect 
of the {\it f}-electron 
superlattices is of great importance and all our previous
statements about interference of the Kondo effect of different KLLs
remain valid.  

Furthermore, we observe that the local occupation of conduction
electrons depends on the layer, see Fig. \ref{dopeAAB}. 
Thus the system forms a charge density wave corresponding to the
superlattice structure.
This corresponds to the fact that different
materials or different microscopic models react differently to a
change of the chemical potential. 
Due to the local singlet formation between conduction electrons and
localized spins in the KLLs, which energetically favors a state with
one conduction electron per lattice site, we find that the occupation
number of the KLLs is  closer to one than the occupation of
the NILs.
However, as the spatial symmetry along the
superlattice structure is broken from the beginning by constructing
the superlattice, there occurs no phase transition, but the charge
density wave corresponds to the superlattice structure.

\section{Conclusion}

We have calculated properties of {\it f}-electron
superlattices. We have elucidated the question, what happens to the
full gap of the isolated Kondo lattice layer, when it is inserted into
a metallic host. We have shown that due to tunneling of
electrons from the metallic layer into the Kondo lattice,
the full gap of the isolated KLL is changed into a pseudo gap.
However, by coupling the KLLs to the NILs, not only the DOS of the KLL
is changed. We have demonstrated that the Kondo
effect of the {\it f}-electrons also has strong
influence on the non-interacting layers, which show pronounced peaks
or dips at the Fermi 
energy. 
Remarkably, these peaks and dips in the NILs are
strongly influenced by interference between the {\it f}-electron
layers. While the configurations with an odd number of NILs 
show a constructive interference, which enhances the resonance at
the Fermi energy, an even number of NILs results in a destructive
interference, canceling all visible effects of the Kondo effect on the
DOS at the Fermi energy in the (1,2)-superlattice.
The Kondo effect also strongly affects the conductivity at low
temperatures.
Because of the formation of the Kondo gap in the KLLs, the superlattice
becomes strongly anisotropic at low temperatures. 
Decreasing the temperature below the Kondo temperature, the
out-of-plane conductivity vanishes, while the in-plane
conductivity stays metallic.

In this study, we have shown that the Kondo effect plays an important
role in the superlattice, not only in the {\it f}-electron layers, but
also in the NILs. The possibility to tune the physical properties
such as the behavior of the Kondo effect is one of the advantages of
a superlattice. So far, we have only analyzed the
  paramagnetic state. 
If magnetically ordered states are analyzed, besides
  the Kondo effect the RKKY interaction will also become important.
For weak coupling strengths, the RKKY interaction will be stronger
than the paramagnetic Kondo screening, which will result in a
magnetically ordered state. However, the heavy fermion
materials which 
have been used in the experiments, i.e. CeIn$_3$ and CeCoIn$_5$, 
show large regions in the pressure-temperature phase diagram, where
the paramagnetic 
heavy-fermion state is stable.
Our results, which we have shown here, can be expected to be
valid for this paramagnetic heavy fermion phase, where the Kondo
screening is stronger than the RKKY interaction. 
Because the Kondo effect in competition with the
RKKY interaction plays an important role in {\it f}-electron
materials, we expect that the tunability of the superlattice will
result in intriguing phase diagrams including magnetic phases, which
depend on the structure of the superlattice.
Including magnetic phases into our studies is left as a future project.

We have focused here on artificially created {\it
    f}-electron superlattices. However, there are many naturally occurring
layered {\it f}-electron materials,  e.g. CeCoIn$_5$. 
The described consequences of the Kondo effect on
  the system, i.e. resonances in the DOS at the Fermi energy,  can be
  expected to also play an important role in these compounds. This
  is currently under investigation.

\begin{acknowledgments}
RP and NK thank the Japan Society for the Promotion of Science (JSPS)
for the support  through its FIRST Program. 
NK acknowledges support through KAKENHI (No. 22103005,
No. 25400366). YT is supported through KAKENHI (No. 23840009).
The numerical 
calculations were performed at the ISSP 
in Tokyo and on the SR16000 at  YITP in Kyoto University.
\end{acknowledgments}


\end{document}